%% file: probe-streaming-arxiv.tex
\documentclass[sigplan]{acmart} 
\renewcommand\footnotetextcopyrightpermission[1]{}

\acmConference[arXiv.org]{arXiv.org}{2021}{}
\acmYear{2021}
\acmISBN{}
\acmDOI{}
\startPage{1}

\setcopyright{none}

\usepackage[nobiblatex]{xurl} 

\newcommand{\etal}{et al.~}

\newcommand{\irradiance}{color}
\newcommand{\Irradiance}{Color}
\newcommand{\visibility}{visibility}
\newcommand{\Visibility}{Visibility}

\usepackage{textcomp}

\usepackage{url}
\usepackage{soul} 
\usepackage{balance } 
\usepackage{tabularx} 
\usepackage{multirow}
\usepackage{mwe}
\usepackage{makecell}
\usepackage{float}
\usepackage{array,graphicx}
\usepackage{enumitem}
\usepackage{pdfpages}

\newcommand{\STAB}[1]{\begin{tabular}{@{}c@{}}#1\end{tabular}}

\usepackage[T1]{fontenc}

\begin{document}
\title[Distributed, Decoupled  Losslessly Light Probe Streaming]{A Distributed, Decoupled System for Losslessly Streaming Dynamic Light Probes to Thin Clients}
\author{Michael Stengel}
\affiliation{\institution{NVIDIA}}
\email{mstengel@nvidia.com}

\author{Zander Majercik}
\affiliation{\institution{NVIDIA}}
\email{amajercik@nvidia.com}

\author{Benjamin Boudaoud}
\affiliation{\institution{NVIDIA}}
\email{bboudaoud@nvidia.com}

\author{Morgan McGuire}
\affiliation{\institution{NVIDIA}}
\email{mcguire@nvidia.com}

\input{sections/abstract}


\begin{teaserfigure}
  \includegraphics[width=\textwidth]{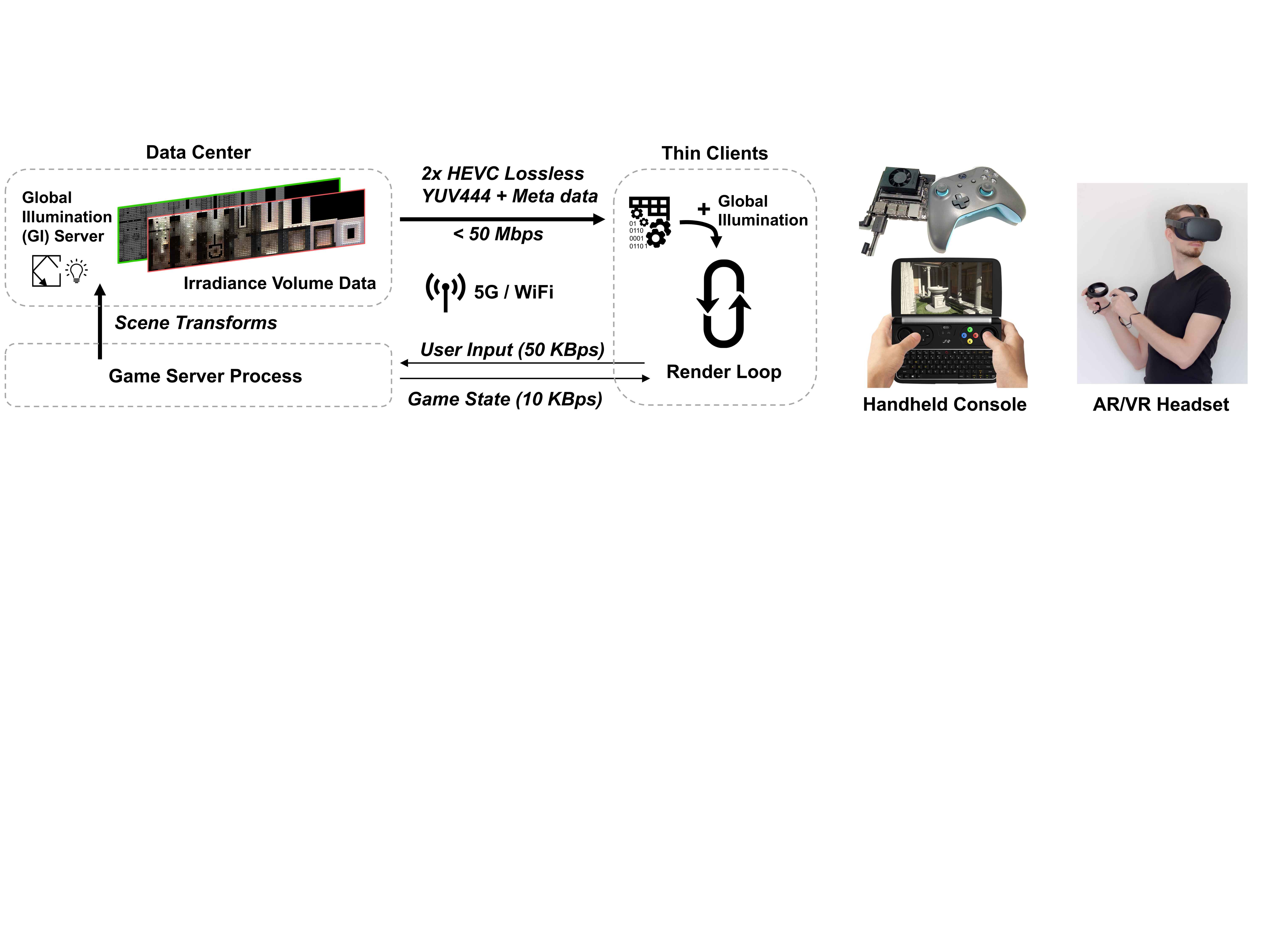}
  \caption{\textbf{Light probe streaming architecture.} Our system streams irradiance volume data from a cloud server to thin clients for mobile gaming and AR/VR. The streamed data enables dynamic high quality ray traced indirect lighting on the clients at low computational cost. We use efficient encoding and decoding hardware features to achieve high compression ratios with low latency. Thus a single server can update thousands of light probes per seconds across multiple connected thin clients, amortizing rendering costs.}
  \label{fig:teaser}
  \vspace{0.5cm}
\end{teaserfigure}

\maketitle

\input{sections/structure}

\bibliography{probestreaming}
\end{document}

%% file: sections/abstract.tex
\begin{abstract}
We present a networked, high-performance graphics system that combines dynamic, high-quality, ray traced global illumination computed on a server with direct illumination and primary visibility computed on a client.This approach provides many of the image quality benefits of real-time ray tracing on low-power and legacy hardware, while maintaining a low latency response and mobile form factor.

As opposed to streaming full frames from rendering servers to end clients, our system distributes the graphics pipeline over a network by computing diffuse global illumination on a remote machine. Diffuse global illumination is computed using a recent irradiance volume representation combined with a new lossless, HEVC-based, hardware-accelerated encoding, and a perceptually-motivated update scheme.

Our experimental implementation streams thousands of irradiance probes per second and requires less than 50 Mbps of throughput, reducing the consumed bandwidth by 99.4\% when streaming at 60 Hz compared to traditional lossless texture compression.

The bandwidth reduction achieved with our approach allows higher quality and lower latency graphics than state-of-the-art remote rendering via video streaming. In addition, our split-rendering solution decouples remote computation from local rendering and so does not limit local display update rate or display resolution.
\end{abstract}

%% file: sections/structure.tex
\input{sections/introduction}
\input{sections/relatedwork}
\input{sections/system}
\input{sections/results}
\input{sections/discussion.tex}

%% file: sections/introduction.tex
\section{Introduction}\label{sec:introduction}

\paragraph*{Motivation}
Today's high performance graphics systems approach cinematic rendering quality in real time using ray traced global illumination, with interaction latency measured in milliseconds. For applications like virtual and augmented reality (VR/AR), the need for low latency and high-fidelity rendering is augmented by a need for a small form factor and low thermal requirements to facilitate user comfort.

A small form factor and good thermals, however, directly constrain high-fidelity rendering at low latency. This constraint becomes more severe as rendering requirements increase and desired form factor and thermal budget decrease, making it impossible to render high quality graphics directly on AR/VR/mobile devices.

\paragraph*{Distributed Graphics}
The state of the art for accelerating high-quality rendering on heterogeneous, low-end devices (AR/VR/mobile) is interactive video streaming. Several cloud gaming services distribute the graphics workload by running games on remote servers. These systems stream user inputs to the server and rendered video back to local clients. At standard dynamic range and 60 Hz frame rate, these solutions already consume moderate bandwidth despite leveraging high efficiency video encoding (HEVC). They can also experience significant latency, as the display framerate is limited by the full network round trip time. In addition, rendering each user view as a full frame does not amortize computation over multiple users, causing the graphics workload to scale linearly with the number of users.

\paragraph*{Decoupled Global Illumination}
The computational cost of streaming full frames scales linearly because of view-dependent rendered effects: each client has a different viewpoint and so a completely different image must be rendered. High-fidelity rendering, however, is composed of multiple simulated effects, not all of which are view dependent. This suggests separating the graphics pipeline into view-dependent and view-independent effects and rendering the latter on a remote machine. This is our approach.

We observe that, unlike direct illumination, \emph{diffuse} global illumination is view independent. We therefore compute diffuse global illumination on a remote machine and stream that data to a thin client where it can be efficiently queried during local rendering.
Shifting the diffuse global illumination (also referred to in our paper as global lighting) off-device has several advantages. As with full frame streaming, it avoids the hardware limitations of thin clients by rendering state-of-the-art ray-traced global illumination remotely, and it provides natural access to new hardware features that can be integrated in a server long before they are propagated to thin clients.

This approach also has the unique advantages that (a) data for global illumination can be amortized over multiple users in the same virtual environment because it is not view dependent; decoupling it from users, frames, and pixels allows increased scaling and lowers the total cost of rendering, 
(b) all but the crudest approximations of global illumination are too expensive to compute on low-end graphics devices, so computing global illumination remotely increases quality by enabling higher fidelity approximations, and (c) significant lag in diffuse global illumination updates is perceptually tolerable (up to nearly half a second)~\cite{Crassin2015CloudLight}.

\paragraph*{Contributions}
To the best of our knowledge our paper presents the first distributed system that enables streaming low-latency, dynamic, high-quality, ray-traced global illumination with visibility information from a remote machine to one or more thin clients.
Our contributions are:
 \begin{enumerate}
     \item A system design for multi-client, distributed, high-quality update and streaming of irradiance volumes with visibility data
     \item A low-latency, GPU-accelerated, lossless texture encoding/decoding scheme supporting 3x10-bit color and 2x16-bit visibility information
     \item Per-client prioritization schemes for GI probe updates reducing network traffic and encoding/decoding latency, and amortizing rendering across clients
     \item Specific best practices for avoiding CPU, GPU, or network stalls on servers and clients during lighting and state synchronization
     \item Evaluation on low-end, mobile gaming and AR/VR devices
 \end{enumerate}

%% file: sections/relatedwork.tex
\section{Related Work}\label{sec:relatedwork}

The stages of an application's user interaction loop are: user input \textrightarrow{} simulation \textrightarrow{} CPU draw call generation \textrightarrow{} GPU graphics pipeline (indirect lighting \textrightarrow{} shadows \textrightarrow{} geometric culling \textrightarrow{} primary visibility \textrightarrow{} direct lighting) \textrightarrow{} framebuffer \textrightarrow{} display.
Many of the system design challenges in distributed graphics stem from choosing the most efficient location(s) in this loop for a network connection and creating efficient representations for synchronizing data across it. The goals are to achieve low latency and low bandwidth consumption at high image quality, low system complexity, and relatively low client requirements. 

The two extreme cases, fully local and remote rendering, are widely deployed due to their practical simplicity, while networking between CPU and GPU has been used in limited contexts (e.g., 2D, data visualization). \textit{Split}, also called \textit{hybrid} or \textit{collaborative, rendering}, which uses both client and server-side 3D rendering, is an active research area that includes our work. Split renderers divide the  graphics pipeline into two or more pieces connected by a network. The cost is increased application development overhead, but the opportunity is improvement in nearly all runtime metrics.

The disadvantages are that image quality is gated by the client device, and it is hard to support heterogeneous clients as new rendering features must be implemented for each one separately, sometimes using different algorithms.

\subsection{Streaming Framebuffer}
Remote rendering with a thin client distributes work by sending user input from a client to a server, which renders complete frames and streams them back to the client as video. This is comparable to replacing the connection between user input device and simulation engine, as well as framebuffer and display with a network~\cite{kirchner2003scalable,richardson2011remote}.
This method is used both by cloud gaming services and productivity-based remote desktop applications \cite{di2020network, cai2014toward}.
Typically, for cloud gaming the application runs in a virtual machine (VM) on the cloud or edge server allowing the user input and network-attached framebuffer to appear local to the game engine~\cite{shea2013cloud}.

The advantages of streaming full or partial frames include that it is a lowest common denominator approach suitable for heterogeneous, low-powered clients and leverages the mature technology ecosystem around efficient video compression.
The disadvantages are that end-to-end video streaming can introduce latencies of above 100 ms \cite{di2020network, tolia2006quantifying,choy2012brewing}, which can be hidden for video applications by buffering, but can be unacceptable for interactive applications and competitive first-person video games~\cite{Bierton2012,shea2013cloud,slivar2014empirical,Chen2019Residual,Morgan2019}.
While video compression scales sublinearly in terms of bandwidth~\cite{sullivan2012overview}, the throughput requirements still grow with resolution and frame rate, and image quality can be suboptimal due to lossy encoding. Including additional information unique to 3D rendered content, such as depth and motion vectors, can provide better robustness on unreliable networks~\cite{pajak2011scalable}. Gaze tracking information captured by the client in real time can be used to control block-wise video compression on the server, reducing the consumed network bandwidth up to 50\% for high resolution video \cite{kamarainen2018cloudvr,illahi2020cloud}.

More aggressive stream compression research renders a local, low-quality estimate of each frame, so only differences between the local render and the remote high-quality frame must be streamed~\cite{Levoy1995Polygon,Chen2019Residual}. This is a more efficient compression mechanism than na\"{\i}ve video, but it does not address the latency or scaling problems of producing per-frame, per-pixel results on a server~\cite{Chen2019Residual}.

\subsection{Streaming Shaded Textures}
In this approach, texture space shading is used to reorder the graphics pipeline with direct lighting before visibility, and the system is split at this point, decoupling shading from local refresh rate and resolution~ \cite{hillesland2016texel}.

The systems challenges include computing and streaming dynamic atlases of texture-space lighting data while avoiding overshading (rendering results at too high a resolution or for unseen areas of the texture). The algorithmic challenge is that most texture-space solutions are restricted to Lambertian surfaces and cannot produce effects such as as glossy highlights.
All previous systems implementing these solutions \cite{Mueller2019AtlasStreaming, Hladky2019tessellated}  have either no direct shadows or high latency in the shadows, which is why we focus on indirect lighting and keep direct visibility shading local to the client.

Mueller \etal~\shortcite{Mueller2019AtlasStreaming} achieve end-to-end latency of 86 ms for lossy results at 40 Mbps. Hladky \etal \shortcite{Hladky2019tessellated} use fewer samples due to better atlas packing but require a bandwidth of 45 Mbps. With comparable network performance, our system solves a complementary problem by providing lossless indirect lighting with both diffuse and rough glossy reflections.
The open problem that remains is combining or extending the described solutions to solve the problem of low latency, remote, direct glossy reflections.

\subsection{Streaming Indirect Lighting}
Computing view-independent indirect lighting on the remote server and primary visibility on the local client is another interesting approach. This has the advantage of providing minimal latency for interaction and camera movement, while offloading the most expensive and hardware feature-driven aspect of the rendering pipeline to the cloud. Using this approach, indirect lighting updates can be decoupled from frame rate and resolution. The added latency in lighting effects has been shown acceptable for diffuse indirect light, but can be problematic when sharp reflections, shadows, or direct illumination are delayed~\cite{Crassin2015CloudLight}. 

Previous work describes good solutions for asynchronously updating and efficiently streaming irradiance maps, photon maps, voxel lights \cite{Crassin2015CloudLight}, virtual point lights \cite{bugeja2019asynchronous,magro2019interactive}, light propagation volumes~\cite{liu2017lightweight,Liu2018Baking},
and irradiance maps combined with probes~\cite{Zabriskie2018Netlight}.
These methods use a variety of lossy video and general lossless encoding techniques.

We extend this body of split lighting work with a system design that targets the same point in the pipeline with novel lossless video encoding and a newer lighting representation: Dynamic Diffuse Global Illumination (DDGI) light probes~\cite{Majercik2019Irradiance}.
However, our system is equally applicable to other GI techniques using irradiance volumes \shortcite{GregerIrradianceVolume1998,HDRTheoryPractice2006, ScherzerRadianceCaching2012} or voxel grids represented as textures \shortcite{Crassin2015CloudLight}.
Generally, the techniques we develop are applicable to any GI solution amenable to amortization over multiple clients and representation as streamed texture data.

\begin{figure*}[t]
\centering
\includegraphics[width=\textwidth]{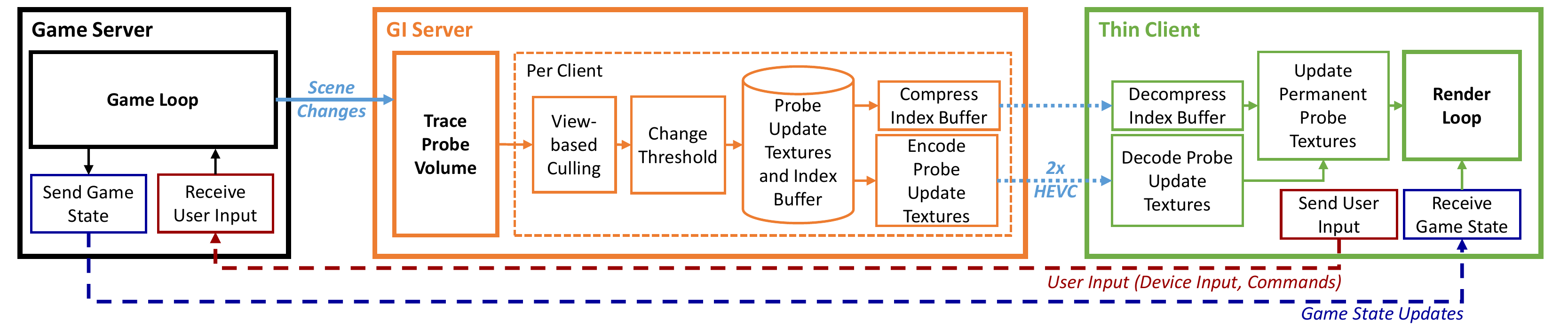}
\vspace{-0.5cm}
\caption{\textbf{Data flow in light probe streaming architecture.}~\normalfont From left to right for a single client: Game simulation loop on game server node (\textit{black box}). Server-side irradiance volume rendering as well as per-client probe selection and encoding on GI server node (\textit{orange box}). Network transfer of data to and from the client (\textit{dashed lines}). Decoding of light probe data and shading on client (\textit{green box}). Note that the game and GI server nodes are only conceptually separated, depending on the number of users and game both can run on a single machine in the data center.}
\label{fig:dataflowarchitecture}
\end{figure*}

\subsection{General Texture Compression}
Fixed-ratio, block texture compression schemes such as DXTn, BCn, or ASTC are often GPU-accelerated for fast random-access decompression \cite{van2007real, nystad2012adaptive}. However, these schemes are optimized for efficient storage of single textures in content creation pipelines such as light maps and therefore are not ideal for real-time encoding and continuous streaming as they do not exploit information of previously streamed texture data \cite{holub2013gpu}.
State-of-the-art video compression algorithms such as H.265 (HEVC) allow significantly better compression ratios through temporal reuse and bidirectional block prediction, and much higher encoding performance provided by dedicated hardware units on the GPU.

With the advent of mobile AR/VR and low-cost volumetric capture devices, video-based point cloud compression and light field streaming techniques have recently received more attention \cite{schwarz2018emerging, wijnants2018standards}. 

Broxton \etal achieve streaming of preprocessed lightfield videos encoded as multi-sphere image representations that allow specular effects and transparencies through view-dependent alpha blending of multiple scene layers \cite{broxton2020immersive}. Despite efficient, lossy mesh compression and HEVC color atlas compression the approach requires high data rates of 124 Mb/s to 322 Mb/s for streaming.

Our implementation is a \textit{sparse} light field scheme resulting in lower consumed bandwidth, as it encodes and streams dynamic, prefiltered \irradiance\ and \visibility\ data for light probes.

Schwarz \etal propose a projection-based compression scheme for 3D scene data. A sequence of point clouds is replaced by a sequence of textures and geometry images after the 3D object is projected to 2D surfaces \cite{schwarz2018emerging}.
Our approach differs as we only stream indirect lighting data and rely on primary visibility being rendered on the client.

Wilson \etal propose the RVL lossless compression scheme to reduce the bandwidth required to transfer 16-bit, single channel depth images \cite{wilson2017fast}. The algorithm provides low compression and decompression times (1 ms on multi-core CPU) and compression ratios of about 3:1. However, since the algorithm compresses individual images the achievable compression ratio is limited compared to video encoding strategies.

%% file: sections/system.tex
\section{System Design}\label{sec:system}

In this section we present our system to stream GI data from a cloud rendering server to thin clients.
The streamed data enables dynamic, high quality, ray traced diffuse GI on thin clients with no ray tracing capabilities, at low computational cost.
We use efficient encoding and decoding hardware features to achieve high compression ratios with low latency.
Further, we present heuristic light probe update schemes motivated by previous perceptual findings
to reduce the amount of data transferred within a GI update.
As a result a single server can update (tens of) thousands of light probes per second, across multiple connected thin clients, amortizing the rendering costs.

\subsection{Architecture Overview}

Our light probe streaming pipeline can be summarized in four steps:
\begin{enumerate}
    \item Server-side ray tracing of the light probe volume
    \item Per-client optimized light probe data encoding
    \item Network transfer of encoded data to the client
    \item Client-side light probe data decoding and rendering
\end{enumerate}

Figure \ref{fig:teaser} demonstrates the high-level design principle, whereas the data flow is described in Figure \ref{fig:dataflowarchitecture}.
Our client-server architecture distributes the lighting pipeline across a server and (multiple) clients instead of streaming full video frames.
The server, assumed to be in a data center, conceptually consists of two nodes interconnected with a high bandwidth link.
The first node acts as the game server to reliably receive user input from clients and update the game state accordingly. This is analogous to a "typical" multiplayer game server.
The second node, the GI server, performs ray tracing of light probe data (Fig. \ref{fig:textures}) and encodes it for client transport.
Any scene change is assumed to be reported to the GI server by the game server with negligible latency.

Using the current scene state, the GI server computes diffuse light probe texture data that is view-independent.
As a result, this GI data can be reused by multiple clients in multiplayer game scenarios, amortizing render costs.

We exploit hardware-accelerated video encoding on the server to compress the light probe data.
The encoded data is transferred to the client(s) over a combination of WiFi, 5G, and/or wired connections using a reliable low latency network protocol. Specifically, our prototype implements reliable UDP provided by the ENet library \cite{enet2020}, although our architecture does not rely on any particular low-latency protocol or network implementation.

On the client side we use hardware video decoding (when available) for decompressing the light probe data with low latency.
The client render loop then uses the uncompressed light probe textures to add dynamic and high quality diffuse GI at low computational cost.
The GI texture decoding process is decoupled from the render loop and can happen at a completely different rate.
This allows the client to run at full frame rate resulting in minimal perceived latency on user input or movement.
Input latency, as it occurs in traditional cloud-rendering solutions, is therefore avoided.

Rendering GI data in the cloud offers specific benefits such as:
(1) diffuse GI is view-independent and used across multiple users and potentially multiple frames;
(2) rendering GI data is computationally too demanding for thin client hardware; and
(3) users are less sensitive to delayed diffuse lighting \cite{Crassin2015CloudLight} than view-dependent effects.

In the remainder of this section we explain our probe data encoding and decoding in detail.
First, as a baseline in Section \ref{sec:fulltexturestreaming} we present streaming of complete and uncompressed irradiance volumes once per GI update.
Second, in Section \ref{sec:texturecompression} we show how low-latency texture compression can significantly reduce the required network bandwidth per update.
Furthermore, in Section \ref{sec:selectiveprobeupdate} our scene-dependent and user-dependent selection strategies reduce the number of transferred light probes.
In combination, our optimizations produce network bandwidth reductions of \textbf{10x to 100x} (depending the amount of changing light probe \irradiance\ and \visibility) in comparison to uncompressed data.

\subsection{Uncompressed Light Probe Streaming}\label{sec:fulltexturestreaming}

\begin{figure}[t!]
\includegraphics[width=\linewidth]{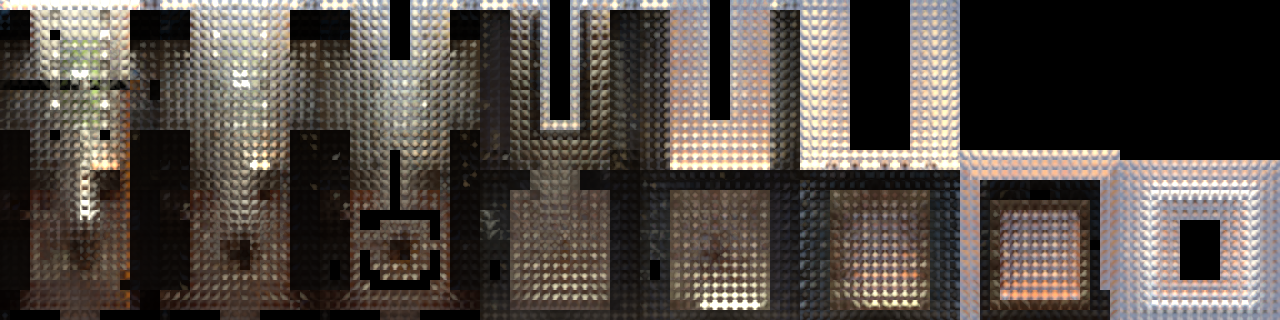}
\vspace{-0.3cm}
\includegraphics[width=\linewidth]{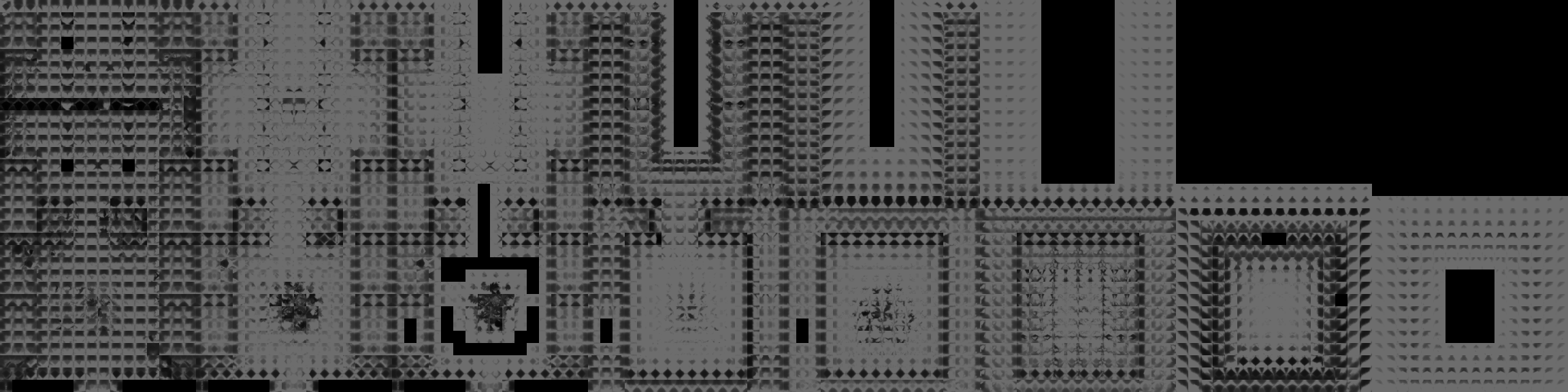}
\caption{\textbf{DDGI light probe volume textures.}~\normalfont Light probe textures for the \textsc{Greek Villa} scene.
The \irradiance\ texture (A2RGB10, top) contains 2048 individual light probes represented by 10x10 texel blocks.
Black pixels represent inactive probes. The \visibility\ texture (RG16F, bottom) includes depth values
in 18x18 blocks per probe (only red color channel shown, values converted to normalized grayscale).}
\label{fig:textures}
\end{figure}

\paragraph*{Streaming Probe \Irradiance}
Our decoupled rendering system adopts the irradiance volume approach by Majercik \etal \cite{Majercik2019Irradiance}.
Each irradiance probe uses a 10x10 texel array, with 32 bits of color per texel, to encode \irradiance\ data for each direction in form of an octahedral mapping (Fig.\ref{fig:textures}, top).
Thus, the required network throughput $T_{\irradiance}$ for updating any given $N_{probes}$ \irradiance\ probes, at a rate $R_{\irradiance}$ (in Hz), is given by Eq.~\ref{eq:IrrThroughput}.
\begin{equation}
T_{\irradiance} = R_{\irradiance} \times N_{probes} \times 3.125\text{ kb}
\label{eq:IrrThroughput}
\end{equation}
A 16x8x16 (2048) probe volume has an uncompressed DDGI \irradiance\ texture size of 6.25 Mb (0.78 MB).
Updating 2048 uncompressed \irradiance\ probes at 10 Hz requires 62.5 Mbps of bandwidth.

\paragraph*{Streaming Probe \Visibility}
DDGI stores mean distance and mean squared-distance for each probe. During shading, these values are used to determine \visibility\ weights between probes and shaded points using a Chebyshev statistical test \shortcite{Majercik2019Irradiance}.
We refer to the mean distance/squared-distance data as "\visibility\ data" and to the texture itself as the "\visibility\ texture" (Fig.\ref{fig:textures}, bottom).

The \visibility\ texture contains 18x18 texels per probe encoded as a pair of half-precision (16-bit) floating-point values (32 bits/texel).
Thus the required throughput ($T_{\visibility}$) for a \visibility\ update rate of $R_{\visibility}$ is given by Eq.~\ref{eq:VisThroughput}.
\begin{equation}
T_{\visibility} = R_{\visibility} \times N_{probes} \times 10.125\text{ kb}
\label{eq:VisThroughput}
\end{equation}

For the same probe volume as analyzed above for \irradiance\ (2048 probes), the size of the \visibility\ texture is 3.24 times higher than that of the \irradiance\ texture, or 20.3 Mb (2.5 MB).
Streaming the raw \visibility\ texture at 10 Hz requires a throughput of 202.5 Mbps.

The minimum required throughput for uncompressed \irradiance\ and \visibility\ textures, updated at 10 Hz for 2048 probes, is ~265 Mbps. This is above practical bandwidth limitations, especially when considering multiple thin-client devices in common wireless networks.
Though this naïve, uncompressed approach is lossless and maintains perfect probe "coherence", it requires unreasonably high network throughput as it sends all probe data regardless of whether the data is unchanged or likely to be used on the client.
In the following sections we describe how this throughput can be reduced.

\subsection{Low-latency Light Probe Compression}\label{sec:texturecompression}

We target GPU-accelerated high dynamic range (HDR) video compression using the High Efficient Video Coding (HEVC) standard \cite{sullivan2012overview}.
In comparison to earlier video compression schemes, such as H.264, HEVC allows for higher compression rates mostly due to better prediction features.

Specifically, our system exploits hardware-accelerated HDR10 encoding and decoding. The HDR10 profile allows for 10-bit color depth in videos whereas earlier low-dynamic range profiles have been limited to 8 bits. In the near future, the HDR10+ profile will provide an option for up to 16 bit color depth through dynamic metadata, though this is not (yet) an established standard and therefore not accelerated in hardware \cite{hdrplus2019hdrplus}.

Hardware-accelerated HEVC decoders supporting HDR video are an important feature being widely adopted in the ecosystem of mobile devices, TVs, and gaming consoles as more HDR content becomes available \cite{flatpanel2019hdr}.
AR/VR platform manufacturers have also announced HEVC decoding support for their next generation of display devices \cite{qualcom2020snap865}.

However, \textit{encoding} HEVC videos is computationally expensive due to the complexity of integrated prediction models.
GPU manufacturers have, only recently, integrated efficient HEVC encoders in hardware, allowing for a dramatic improvement in encoding performance in real time \cite{Nv2020supportmatrix, Nv2020tegrasupportmatrix}.

\paragraph*{Encoding \Irradiance}

Since our goal is maintaining the original quality of ray traced irradiance volumes we implement \textit{lossless}  encoding and compress \irradiance\ data as follows.

\begin{figure}[t!]
\includegraphics[width=\linewidth]{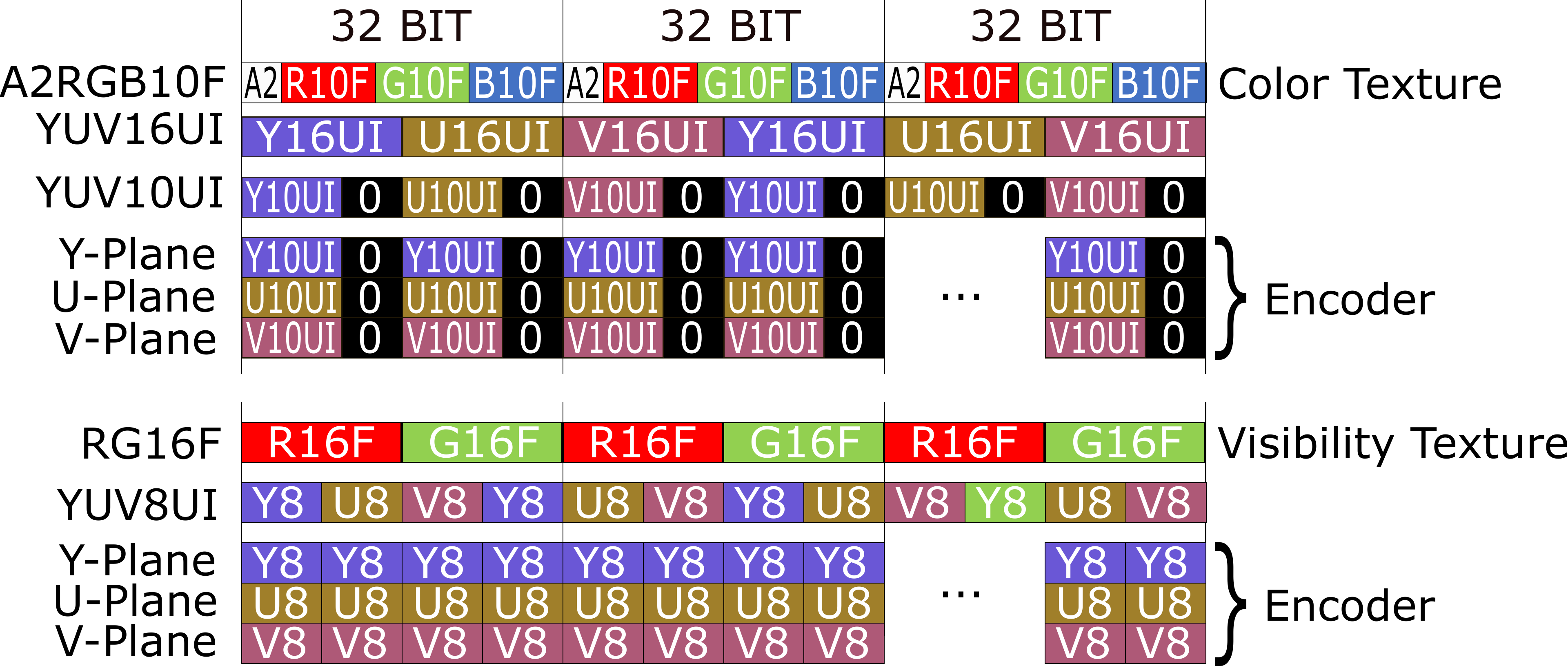}
\caption{\textbf{Memory layout for lossless texture encoding and decoding.}~\normalfont
Top: The normalized values in the A2RGB10 \irradiance\ texture are quantized to 10-bit unsigned integer color values and bit-shifted into 16-bit unsigned integer YUV values.
The original 2-bit alpha values are not used. For hardware encoding the YUV tuples are reordered into YUV planes.
Bottom: The \visibility\ texels representing half-precision floating-point depth values are directly copied into 8-bit YUV planes to avoid quantization loss.
For decoding the steps are performed in reverse order.}
\label{fig:memorylayout}
\vspace{-3mm}
\end{figure}

Normalized \irradiance\ values are saved in a A2RGB10F texture (10 bits unsigned small floats for each color, 2 unused bits for alpha).
As the unsigned small floats are normalized we can quantize these values, without loss, into 10-bit unsigned integer values in the range [0,1023].
Our input format for the hardware encoder is 16 bit unsigned integers organized in 3 Y,U,V planes (linear Y-luma, U,V-chrominance).
We use the YUV444 surface format to avoid chroma subsampling for lossless compression.
The hardware encoder uses only 10 out of 16 bits for \irradiance\ encoding.
The remaining bits are reserved for future versions of the codec which support 12-bit and 16-bit color encoding \cite{hdrplus2019hdrplus}.
Therefore we set these bits to zero (Figure \ref{fig:memorylayout}).
The YUV tuples are then reordered into Y,U,V planes over which hardware encoding is applied.
For decoding the process is performed in reverse.
%

\paragraph*{Encoding \Visibility}
Our solution for \visibility\ leverages \textit{lossless} and \textit{low bit depth} video encoding.
The idea is to distribute a single 16-bit floating-point value across two adjacent 8-bit integer values (Figure \ref{fig:memorylayout}, bottom).
As the \visibility\ texture for a light probe holds two channels (RG16F) we distribute the bits across four 8-bit integer values.
The utilized hardware encoder does not support 4-channel images. As a workaround, we pack a sequence of three RG16F values into a sequence of four YUV values effectively distributing the bits into YUVY, UVYU, VYUV sequences.
We widen the YUV texture to allocate enough memory for the \visibility\ texture.
For an original light probe \visibility\ texture width $x$ we increase the YUV texture width to $x' =  \lceil 4/3 \cdot x \rceil$.

\paragraph*{Threaded Client-side Decoding}
In order to maximize throughput and avoid CPU blocking behavior between rendering updates and decoding incoming encoded GI data, texture decoding happens in a separate thread on the client. This unblocks the game loop during CPU-based texture decoding or GPU decoding dispatch.

\subsection{Selective Light Probe Updates}\label{sec:selectiveprobeupdate}
As a next step in optimizing network throughput per GI update we attempt to avoid sending those probes which do not contain information relevant to client-side shading. A relevant probe satisfies three conditions:
\begin{itemize}
    \item It is being updated and used for shading in the scene
    \item Its affiliated texels have changed significantly since its last transmitted update to a client
    \item It is shading points potentially visible to a user
\end{itemize}
On the server, we gather all probes for which the preceding three conditions hold and transmit them to the client.
At each transmission, we record transmitted probe data on the server and compare it to rendered probe data to determine which probes should be sent in the next transmission.
Details of probe gathering and changed texel computation are given in the supplemental material.

\paragraph*{Light Probe Repacking} Inactive probes can be removed from the original texture by moving to an index-based probe update system (selective, iterative) as opposed to a full texture transfer (global probe volume update).
An additional reduction is possible for each core texel array in both \irradiance\ and \visibility\ texture.
The original DDGI textures include 1-pixel guard bands (per probe) for texture filtering that contain redundant information from the core texel array. This data can be removed in the packing step on the server and reconstructed without loss on the client during unpacking.
The strategy reduces the size of transferred textures resulting in a benefit that is independent of the compression scheme (or uncompressed transfer).
However, for temporal reuse, probe locations in the textures should be maintained over multiple frames if possible.

The index-based update scheme allows us to adjust the total packet size in flexible ways, as the number of iteratively updated probes can be specified.
This effectively introduces a 'knob' to adjust throughput on a per-client basis.

In any given (well-placed) probe volume about 75\% of probes are active \cite{Majercik2019Irradiance}. Thus, by simply excluding probes located within/outside of static geometry a 25\% reduction in network throughput (or 33\% increase in maximum update rate) can be achieved.
Removing the guard bands decreases the \irradiance\ data by 36\% and the \visibility\ data by 21\% per light probe considering a single texture set.
In combination, we expect a reduction of 52\% for \irradiance\ and 40.7\% for \visibility\ resulting in a reduction of 47.2\% overall.
This repacking scheme reduces the light probe update bandwidth by almost half without any loss of information compared to streaming complete irradiance volume textures.

\paragraph*{Light Probe Atlas Texel Arrangement}
As each probe texel represents directional information, it is understandable that adjacent probes contain angular redundancy. We  experimented with different checkerboard patterns of probes with the goal of placing texels of neighboring probes with similar directions (used for probe tracing) adjacently in the probe atlas. However, we observed that the lossless HEVC video encoder is able to exploit redundancy best in the original DDGI layout where individual probes are arranged as self-contained blocks.

As video encoding is performed on blocks in a (large) atlas we reason that the encoder has a better chance in exploiting redundancies across \textit{all} probes in the irradiance volume than from just angular redundancy of \textit{neighboring} probes. Please find details of this analysis in the supplemental material.

\paragraph*{Light Probe Update Textures} In the following we describe our strategy for selective probe updates on the server and client(s).
The data flow for the described processes are visualized in Figure \ref{fig:dataflowarchitecture}.
Again, additional algorithmic details are provided in the supplemental material.

The server stores permanent \textit{client probe textures} as well as temporary selective \textit{probe update textures} for each client. 
The permanent textures always represent the state of probes as they exist on each client.
The temporary probe update textures are created per update step and store the probe information that is being renewed.
Along with this texture, the client needs to be informed of which probes in the volume are being updated.
This information is provided by the \textit{probe index buffer} which maps probe update texture coordinates into the client's probe textures. In comparison to the probe update textures, the size of the probe index buffer (encoded as a single 2-bytes unsigned integer per probe) is very small and is further reduced by delta encoding. We observed average index buffer sizes of <1 kB even for large probe volumes.

For the client this process happens in reverse order. Only those light probes that have been sent to the client are updated in the permanent probe textures. The mapping from the update textures to the permanent light probe textures uses the received and decompressed probe index buffer. The remaining texels are assumed valid from the previous state. As the texture decompression and update step happen asynchronously in a parallel thread, the render loop is not blocked and can continuously run at full frame rate.

Note that until this point we have not covered how probes to be updated in the client view are selected and how we evaluate when a probe has changed.
Conservatively, without loss in quality, we can select all probes that are \textit{active} and therefore potentially contribute to shading of any part of the scene on the client side. 
However, to reduce the number of probes further we present optimizations on the server-side probe selection, specifically \textit{Light Probe Change Thresholding} and \textit{Server-side Client View Light Probe Culling}.

\paragraph*{Light Probe Change Thresholding}
To conservatively estimate all of the probes that could affect the primary client view, we consider all probes that have changed in value since the last time they were transmitted. We maintain two additional textures per connected client on the server, one for \irradiance\ and one for visibility, that store the state of each probe at the last time it was transmitted. At transmission time, we compare each probe to its last transmitted state and only consider it for transfer if its values changed beyond a specified threshold. Although higher thresholds might be perceptually tolerable, we conservatively mark a probe as changed if it is not texel-for-texel identical to its last transmitted state. To make this a useful heuristic (and to improve compression rates) we assume fully-converged \irradiance\ on the server making scene changes the only source of variance in the probe result. Before each probe is transmitted, we write its data to the respective last-transmitted texture.

\paragraph*{Server-side Client View Probe Culling}
Given a viewpoint within a scene, the subset of probe texels that might contribute to a rendered frame are only those which are close to points of primary visibility. This represents a, in some cases substantial, reduction of the overall number of active/updating probes from the full volume.

By limiting updated probes to those which contribute to shading the primary visibility frustum, the number of probes required to be updated can be reduced.
Ideally a game-dependent client view prediction scheme could adjust the number of updated probes. However, such a prediction is strongly coupled to network latency and game content.
Thus, we decide for a strategy that is rather application-independent.
On the server we estimate the potential visible set (PVS) of probes contributing to client shading by rendering a spherical view from the position of the client. Gathering probes from this PVS estimation ensures correctness of the client view under camera rotation (see Figure 3 in the supplemental material).
The supplemental material describes this step in greater detail.


%% file: sections/results.tex
\section{Results}\label{sec:results}

\begin{figure*}[t!]
\setlength\tabcolsep{2pt}%
\begin{tabularx}{\textwidth}{ccc}
   \includegraphics[width=0.33\linewidth, keepaspectratio]{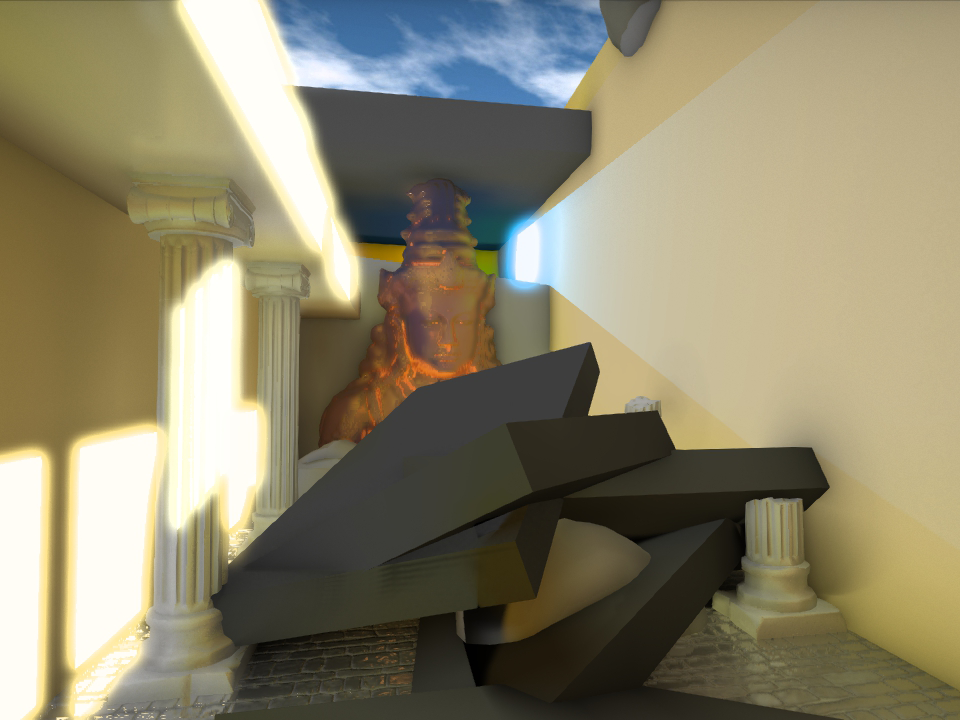} &
   \includegraphics[width=0.33\linewidth, keepaspectratio]{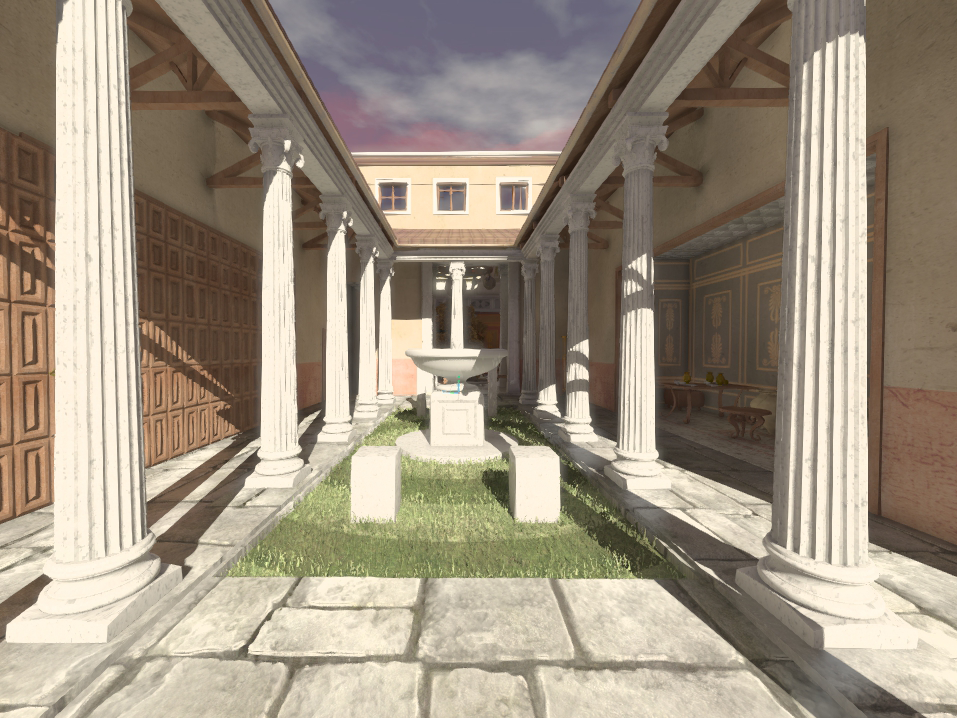} &
   \includegraphics[width=0.33\linewidth, keepaspectratio]{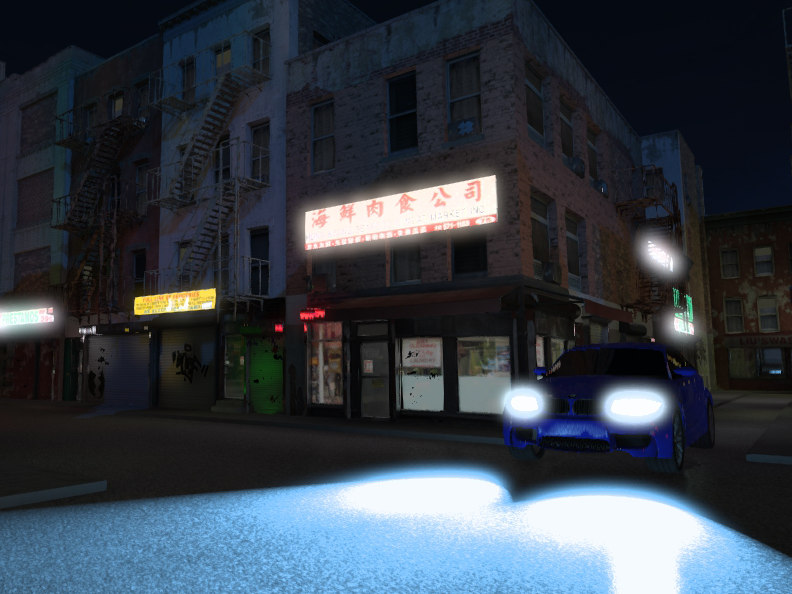}\\
\end{tabularx}

\vspace{0.1cm}

\newcolumntype{R}{>{\raggedleft\arraybackslash}p{0.35\textwidth}}
\newcolumntype{P}{>{\raggedleft\arraybackslash}p{0.2\textwidth}}

\setlength\tabcolsep{2pt}%
\begin{tabularx}{1.0\textwidth}{R|P|P|P}
& \multicolumn{1}{c|}{\textsc{Tomb Buddha}} & \multicolumn{1}{c|}{\textsc{Greek Villa}} & \multicolumn{1}{c}{\textsc{Night Street}} \\
\hline
Scene \# triangles    & 1,195,188         & 4,538,959         & 672,206\\
Drawn \# triangles (with shadow maps)    & 2,391,036         & 9,078,626         & 3,141,330\\
Total GI volume probes  & 2048 (16x8x16)    & 4096 (16x8x32)    & 8192 (32x4x64) \\
Total active probes     & 1580 (77.1\%)     & 3349 (81.7\%)     & 4160 (50.7\%) \\
\end{tabularx}
\caption{\textbf{Scene statistics.}~\normalfont We selected three scenes representative of different games scenarios.
In \textsc{Tomb Buddha} (left) the roof caves in, resulting in highly dynamic geometry and lighting changing from very dark lighting to strong incoming sunlight. The \textsc{Greek Villa} (center) provides always changing lighting conditions due to a fast day/night cycle simulation. The geometry in this scene is mostly static with the exception of an oversized orrery influencing the incoming sunlight in the main room. Finally \textsc{Night Street} (right) contains lighting from a police car and the head lamps of a moving vehicle.
}
\label{fig:sceneoverview}
\end{figure*}

We evaluate our system over three test scenes:
\textsc{Tomb Buddha}, \textsc{Greek Villa}, and \textsc{Night Street} (Figure \ref{fig:sceneoverview}).
Each scene provides different combinations of static/dynamic geometry, lighting conditions, and size. The probe volume is set to span the full scene bounds in each scene.

We evaluate our approach with regard to the network bandwidth required by a client to achieve GI update rates in the range of 10 to 30 Hz. The update rates are chosen to reach perceptually high to very high quality for indirect lighting. Detailed time plots for each scene can be found in the supplemental material. 
Based on previous work by Majercik \etal \cite{Majercik2019Irradiance} we consider DDGI probe grid sizes of up to 16x(8-32)x16 probes (2-8k total probes). We target full volume updates at 10-30 Hz producing probe throughput in the 20k-240k probes/s range. These update rates should allow to always  stay within perceptually acceptable latency limits (< 500ms) for diffuse global illumination \cite{Crassin2015CloudLight}.

We use the NVIDIA Video SDK as the interface to the video encoder on the GPU
that allows to allocate an unrestricted amount of parallel encoding sessions \cite{Nv2020supportmatrix}.
If a client does not support hardware HEVC decoding, we fall back to software decoding with the FFmpeg library at lower performance and higher CPU load.
Prior to encoding we perform bit-wise conversion in a compute shader and copy the result into CUDA memory.
Finally, we make sure to only copy compressed bit streams between CPU and GPU, to minimize bandwidth consumption and memory transfer costs within both the client and the server.

We optimize for latency by setting the encoder frame delay to zero.
We do not encode B-frames as they require future frames.
The remaining intermediate frames only depend on previous frames that have been sent to the client already.

This trades off delay for compression performance, but allows our system to minimize overall latency in encoding and decoding.
We also tune compression using the group of pictures (GOP) parameter for controlling the interval of reference frames. Low values are used for unreliable connections, where lost data requires a new I-frame to recover.
As we use a reliable communication protocol, we use a high GOP length of 30 frames.

\paragraph*{Testbed Systems} For our server-side tests we use a remote workstation (Intel i7-4930K CPU, 28 GB host memory, NVIDIA Quadro RTX 8000 GPU), hosting our minimal game logic and the GI updates in parallel.
This remote server connects to the client via internet (3 hops), an additional Wifi router (ASUS RT-AC5300) and an additional signal repeater on the client side representing a common wireless home network configuration for online gaming and streaming.

For the client we tested 3 different hardware configurations: (1) a current generation gaming GPU (NVIDIA RTX 2080~Ti containing a single hardware video decoder) connected to a VR headset (Oculus Quest), (2) a mobile GPU platform (NVIDIA Xavier NX) representing a next-generation portable gaming console with two hardware decoding units, and (3) a handheld gaming platform (GPD Win 2) with only CPU-based decoding support.

NVIDIA Xavier NX and the GPD Win 2 (Fig.~\ref{fig:teaser}) both have greatly reduced system specifications compared to a gaming GPU \cite{wingpd2020,nvxaviernx2020}. As a result, in order to allow for local rendering of direct visibility the \textsc{Greek Villa} scene was modified to reduce vertex count, resulting in a total triangle count of 786,030 and a total drawn triangle count of 1,572,735. The light probe volume size was also reduced to 16x8x8 (from 16x8x32).

\paragraph*{Compression Performance}
Table \ref{table:compressionanalysis} presents lossless compression ratios for full volume encoding (including inactive probes and probe guard bands).
We show these numbers as upper bounds as they include all probes that can change from frame to frame.
In practice the size of the encoded probe textures are smaller as we provide several additional probe reduction strategies.
The mean compression ratio ($r_{avg}$) is estimated by averaging encoding performance over 1000 frames.
The worst compression performance is typically affiliated with independent reference frames (I-Frames). Higher compression is achieved for predicted frames (P-Frames).

We expect the amount of dynamics in lighting and geometry to have a strong influence on the compression performance.
Therefore it is only partly meaningful to compute a compression ratio over all scenes.
Our selected scenes represent scenarios of different dynamics in terms of \irradiance\ and \visibility. As a result, our strategy of exploiting temporal coherence in GI data shows varying success (Table~\ref{table:compressionanalysis}).

The \textsc{Tomb Buddha} scene is the most dynamic scene as the volume is inside a closed room with a collapsing roof. Over time more and more sunlight enters the area. This scenario not only changes the \irradiance\ received by all light probes but also the observed \visibility. Permanently changing all light probe data represents the worst case for our encoding. As a result the video encoder has limited chances to reuse data from blocks from previous frames due to the required lossless encoding ($r_{avg}=3.4$ for \irradiance\ and $r_{avg}=4.5$ for \visibility). 

The \textsc{Greek Villa}, however, achieves a high mean compression ratio for \visibility\ ($r_{avg}=208.8$) as only parts of the scene geometry such as the orrery are changing.

The simulated day-night cycle results in permanent but slower change in indirect lighting than for the \textsc{Tomb Buddha} scene resulting in significantly higher \irradiance\ compression performance ($r_{avg}=14.1$).

The \textsc{Night Street} scene is fairly static overall and as a result we measure very good compression ratios for both textures ($r_{avg}=92.4$ for \irradiance\ and $r_{avg}=1963.4$ for \visibility). The high amount of temporal reuse is close to a best-case scenario for video encoding.

\begin{figure*}[ht]
\centering
\includegraphics[width=\textwidth]{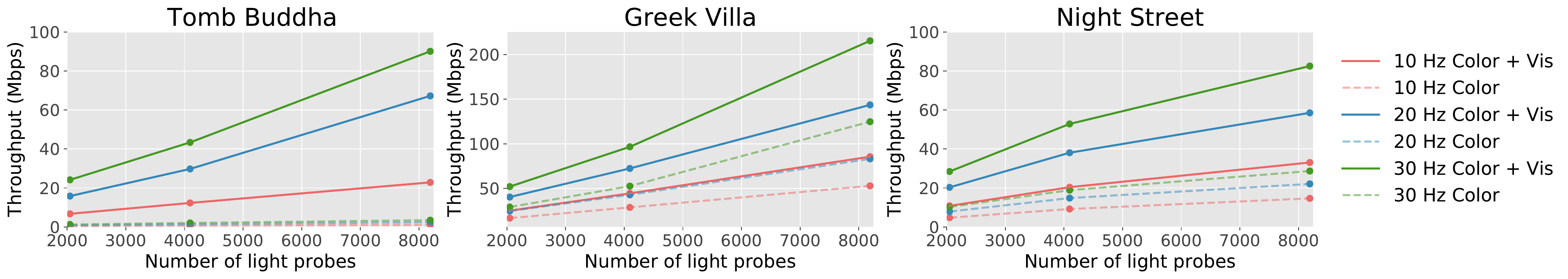}
\caption{\textbf{Consumed network bandwidth for full volume encoding.}~ \normalfont We show results for encoding the full light field volume as described in Section \ref{sec:texturecompression}. For each scene we present the average network throughput for 3 different update rates (10, 20, and 30 Hz). Overall \irradiance\ textures (\textit{Color}) achieve higher compression ratios and contribute less to the overall consumed bandwidth than visibility data (\textit{Vis}). For scenes with static geometry the visibility data would not contribute to the long-term bandwidth consumption.}
\label{fig:throughputfullvolumecompression}
\end{figure*}

These results show that for DDGI data, average video compression not only outperforms
lossless compression routines (2:1) \cite{collet2013lz4} but often even lossy fixed-ratio texture compression techniques (4:1) \cite{nystad2012adaptive} that do not offer to exploit temporal redundancy.
In terms of compression performance, our method is comparable to (lossy) supercompression techniques such as Crunch DXT or Basis \cite{crunch2017} but our technique is orders of magnitude faster.
In addition, exploiting dedicated encoding and decoding hardware units unblocks parallel processes running on the GPU.
We show in Figure \ref{fig:throughputfullvolumecompression} how our solution scales with size of the GI irradiance volume.

For a single frame, our proposed strategy is not ideal in terms of compression ratio ($r_{max}=1.3$ for \irradiance\ and $r_{max}=1.1$ for \visibility\ in the worst cases)
when compared to complex compression schemes that achieve compression rates of 1.5x - 4x in average for floating-point data \cite{lindstrom2014fixed}.
However, the video hardware decoder is optimized to exploit \textit{temporal coherence} in data which occurs frequently in GI data \cite{Crassin2015CloudLight, Majercik2019Irradiance}.
Therefore, when encoding multiple frames video compression rates quickly outperform single-frame fixed-rate compression schemes.
This even holds for our most dynamic scene \textsc{Tomb Buddha} which has permanently changing \irradiance\ and \visibility\ over a high number of probes.

In spite of our efficient compression scheme, 16-bit floating-point data (used for \visibility\ representation) consumes more bandwidth and can be a challenge for large scenes with lots of moving geometry and high probe count. However, we argue that \visibility\ changes are less likely than \irradiance\ changes from dynamic lights, resulting in less frequent  \visibility\ updates.

\paragraph*{Transcoding Latency}
We break down encoding and decoding delays for each scene (Table ~\ref{table:decodinganalysisperscene}) and analyze end-to-end latency (Table~\ref{table:latencybreakdownallprobes}).
Note that the encoding time naturally increases with texture size.
Most importantly however, encoding time depends on the resulting compression performance.

Starting with the smallest scene (\textsc{Tomb Buddha} with 2048 probes) we see an average encoding time increase of 53.9\% for \textsc{Greek Villa} where the number of probes is doubled.
When again doubling the number of probes for \textsc{Night Street} the average encoding time is only increased by 9.7\% over \textsc{Greek Villa}. For \visibility\ we see very different growths, specifically 27.9\% and 60.79\%.
The encoding time increases sub-linearly with texture size as larger probe textures are expected to contain more redundancy.

\begin{table}[t]
\small
\begin{tabularx}{\columnwidth}{p{0.01\columnwidth}p{0.01\columnwidth}p{0.23\columnwidth}ccc}
\cline{1-6}
  &&& \textsc{Tomb} & \textsc{Greek} & \textsc{Night}\\
  &&& \textsc{Buddha} & \textsc{Villa} & \textsc{Street}\\  
\cline{1-6}
\multirow{4}{*}{\STAB{\rotatebox[origin=c]{90}{\textbf{Encoding}}}}
&\multirow{2}{*}{\STAB{\rotatebox[origin=c]{90}{\textbf{Irr.}}}}
& Mean [$\sigma$] & \textbf{1.67} [0.43] & \textbf{2.57} [0.92] & \textbf{2.82} [0.85] \\
&& Min. / Max. &  1.36 / 4.14 & 1.60 / 5.89 & 2.26 / 7.94 \\
\cline{2-6}
&\multirow{2}{*}{\STAB{\rotatebox[origin=c]{90}{\textbf{Vis.}}}}
& Mean [$\sigma$] &  \textbf{3.15} [1.79] &  \textbf{4.03} [2.93] & \textbf{6.48} [3.96] \\
&& Min. / Max. &  2.38 / 13.00     &  3.09 / 19.96     & 5.32 / 27.77 \\
\cline{1-6}
\multirow{4}{*}{\STAB{\rotatebox[origin=c]{90}{\textbf{Decoding}}}}
&\multirow{2}{*}{\STAB{\rotatebox[origin=c]{90}{\textbf{Irr.}}}}
& Mean [$\sigma$] & \textbf{1.40} [0.55] & \textbf{2.62} [0.82] & \textbf{1.82} [1.11] \\
&& Min. / Max. & 0.88 / 4.19 & 1.37 / 6.40 & 1.38 / 7.94 \\
\cline{2-6}
&\multirow{2}{*}{\STAB{\rotatebox[origin=c]{90}{\textbf{Vis.}}}}
& Mean [$\sigma$] & \textbf{2.35} [2.01] & \textbf{2.62} [3.63] & \textbf{4.04} [5.01] \\
&& Min. / Max. &  1.25 / 12.93 &  1.72 / 22.71 & 2.89 / 30.99 \\
\cline{1-6}
\end{tabularx}
\vspace{0.1cm}
\caption{\textbf{Encoding/Decoding performance.}~\normalfont Times for irradiance (Irr.) and visibility (Vis.) encoding and decoding given in milliseconds include data copy between GPU and host memory. Measured on desktop GPU (NVIDIA 2080 Ti).}
\label{table:decodinganalysisperscene}
\vspace{-0.5cm}
\end{table}

To understand these results we have to consider Table \ref{table:compressionanalysis}. The \textsc{Night Street} scene has much higher compression performance due to its rather static nature. Our measured encoding time also includes the required copy from GPU to host memory so that the encoded data can be sent via network.
The copy step is faster if the data compression is more successful.
In addition the time the encoder spends on processing a frame not only depends on the frame resolution and bit depth but also on the chance for temporal reuse. Higher temporal reuse allows for faster processing.

Looking at the longest encoding delay instead of the mean is an alternative way for evaluation.
This happens when an I-frame is encoded and temporal reuse is therefore impossible.
All scenes have comparable worst-case compression ratios ($r_{min}$). When comparing encoding times for these cases we get a more objective analysis of how the encoding time behaves for various number of probes.
For the \irradiance\ texture we observe growths of 42.3\% and 34.8\% when the number of light probes doubles. We measure a comparable increase for \visibility\ data (53.5\% and 39.1\%).

In decoding times, we see comparable values (Table~\ref{table:decodinganalysisperscene} and~\ref{table:latencybreakdownallprobes}, Client A). In particular, the mentioned worst-case decoding timings are very close to their encoding counterparts. For highly compressed frames decoding happens very quickly. As for encoding times, latency reaches its highest values in cases of low compression ratios (I-frames) primarily due to the significantly higher amount of time spent on \textit{uploading} the data to the GPU.
On mobile hardware the copy is avoided due to unified memory (shared between CPU and GPU) leading to fast overall decoding times despite slower decoding hardware (Table \ref{table:latencybreakdownallprobes}, Client B).

The results also show that decoding \visibility\ data is more expensive than \irradiance\ as the data is 3.46x larger (more texels per light probe, higher bit depth) and floating-point numbers are expected to have lower compression performance.

\begin{table}[t]
\small
\begin{tabularx}{\columnwidth}{p{0.02\columnwidth}p{0.4\columnwidth}ccc}
\hline
& & \textsc{Tomb} & \textsc{Greek} & \textsc{Night}\\
& & \textsc{Buddha} & \textsc{Villa} & \textsc{Street}\\  
\hline
\multirow{4}{*}{\STAB{\rotatebox[origin=c]{90}{\textbf{Color}}}}
& Uncompressed size (MB) & 0.78 & 1.56 & 3.13  \\
& Mean Comp. Ratio $r_{avg}$ & 3.4 & 14.1 & 92.4 \\
& Min Comp. Ratio $r_{min}$ & 1.3 & 1.4 & 2.1 \\
& Max Comp. Ratio $r_{max}$ & 4.8 & 16.8 & 164.6 \\
\hline
\multirow{4}{*}{\STAB{\rotatebox[origin=c]{90}{\textbf{Visibility}}}}
& Uncompressed size (MB) & 2.53 & 5.06 & 10.13 \\
& Mean Compr. Ratio $r_{avg}$ & 4.5 & 208.8 & 1963.4 \\
& Min Comp. Ratio $r_{min}$ & 1.1 & 1.1 & 1.5 \\
& Max Comp. Ratio $r_{max}$ & 7.4 & 235.6 & 2807.9 \\
\end{tabularx}
\vspace{0.1cm}
\caption{\textbf{Compression ratios for full-frame lossless encoding.}~\normalfont
The average compression ratio largely depends on the temporal reuse of previous results. Areas of low change in scene geometry and lighting (\textsc{Night Street}) allow for high compression ratios whereas highly dynamic scenes result in lower compression ratios (\textsc{Tomb Buddha}). Minimal compression ratios are given for I-frames. Highest ratios are achieved for P-frames.}
\label{table:compressionanalysis}
\vspace{-0.6cm}
\end{table}

\paragraph*{Full Pipeline Latency}

We show results for the complete GI volume compression pipeline in Table \ref{table:latencybreakdownallprobes}.
The given pipeline includes full probe volume updates and rendering, spanning the time spent on the server, network transport, and client.
For each pipeline stage we use the average result across all three test scenes (from 
Table~\ref{table:decodinganalysisperscene}).

For clarity, the texture encoding step on the server includes both textures (\irradiance\ and \visibility) in sequence.
This strategy ensures coherence of the texture information but represents the worst case in terms of latency.
In practice, these textures are updated at different rates for lower throughput and lower latency and can be encoded and transferred in parallel.

Note that on the server we are using more samples for tracing each probe than proposed in the original DDGI paper by Majercik \etal \cite{Majercik2019Irradiance} (256 instead of 64) as we assume converged probe textures for high image quality and better video compression performance.

On the tested hardware, our scheme is very fast on average considering the fact that all textures are encoded losslessly. The approach is especially beneficial for smaller textures. To be thorough, we also tested lossy compression for \irradiance\ textures resulting as expected in lower decoding times. However, lossy compression requires careful perceptual quality analysis which we leave for future work.
Our analysis shows that steps of the pipeline that we have control over have relatively low latency, with an average total of 22 ms on a powerful client and 34 ms on a thin client using lossless compression (19 and 26 ms when \irradiance\ is lossy compressed).

Depending on the hardware, the copy between GPU and host memory can become the latency bottleneck. However, for the tested probe volume sizes the proposed approach stays well within reasonable bounds for diffuse GI data \cite{Crassin2015CloudLight}.

For expressing the full end-to-end latency we introduce two variables $L_{Tex}$ and $L_{Input}$ describing the network latency for the texture transfer from server to client and the user input data from client to game server. These delays are variable as they depend on the underlying network connection.
We can safely assume 20 ms as a realistic user input delay ($L_{Input}$) and 100 ms as texture transfer delay ($L_{Tex}$). Under realistic internet streaming conditions (with 3 hops) we measure the following average texture transfer latency: 25.2 ms ($\sigma$~=~18.23 ms) for \textsc{Tomb Buddha}, 35.2 ms ($\sigma$~=~13.81 ms) for \textsc{Night Street} and 43.1 ms ($\sigma$~=~24.45 ms) for \textsc{Greek Villa}. Time series plots for latency and encoded texture size for each scene are provided in the supplemental material.
Given these assumptions, the end-to-end latency of 150 ms stays well below the recommended bounds of 500 ms for diffuse GI \cite{Crassin2015CloudLight}.
This leaves a significant headroom of 350 ms for latency 'hitches' that (potentially) occur due to network congestion.

\paragraph*{Selective Probe Transmission}
On the server, all probes are necessary to encode the view-independent irradiance field with visibility. 
Only a fraction of these probes are necessary to correctly shade the client view (see Section \ref{sec:selectiveprobeupdate}).
Culling unnecessary probes allows us to send fewer probes during transmission, thus reducing bandwidth.

In each transmission, we consider only probes that changed since they were last sent. We further cull from this set probes that do not shade any surface potentially visible from the client's camera position. The ratio of probes transmitted using this technique relative to sending all active probes is given in Figure~\ref{fig:SelectiveProbeUpdatePerformance}. Further details of the probe selection process are given in the supplemental material.

\begin{table}[t]
\small
\begin{tabularx}{\columnwidth}{lrp{0.36\columnwidth}ccc}
\hline
~& ~ & Pipeline Stage & \multicolumn{3}{c}{Duration (ms)}\\
 ~&~&~&Server&Client A&Client B\vspace{-0.25cm}\\
\hline
\multirow{3}{*}{\STAB{\rotatebox[origin=c]{90}{\textbf{Server}}}}
& &  &  &  & \\[-0.6em]
&1  & Ray trace probe textures & 5.63 &  & \\
&2  & Texture encoding & 6.91 &  & \\
& & & & & \\[-1em]
\textbf{---} &3  & Network transport && \multicolumn{2}{c}{$L_{Tex}$} \\
\multirow{5}{*}{\STAB{\rotatebox[origin=l]{90}{\textbf{Client}}}}
 &4  & Texture decoding & & & \\
 &  & \hspace{2em} lossless & & 4.95 & 11.34 \\
 &  & \hspace{2em} (lossy) & & (1.63) & (2.78) \\
 &5  & Render frame & & 4.48 & 10.52 \\
 &6  & Client input transfer && \multicolumn{2}{c}{$L_{Input}$} \\
 & & & & & \\
\hline
 \multicolumn{4}{l}{Total time \hspace{2em} lossless ~~ $L_{Tex}$ + $L_{Input}$ +}& 21.97 & 34.40 \\
  \multicolumn{4}{l}{\begin{tabular}{l} \hspace{6em} (lossy)\end{tabular}}& (18.65) & (25.84) \\
   \hline
\end{tabularx}
\vspace{0.1cm}
\caption{\textbf{Latency analysis for full probe volume update.}~\normalfont The given pipeline includes a full probe volume update. For each pipeline step we use the average result across all three test scenes (see results in Table~\ref{table:decodinganalysisperscene}).
Client A measured on workstation GPU (NVIDIA RTX 2080 Ti, decoding in sequence using the single decoder unit), Client~B measured on mobile GPU (NVIDIA Xavier NX, decoding using both decoders in parallel). Timings for lossy \irradiance\ compression are given for reference in brackets. Time series plots and statistics for network traffic and transport latency are provided for each tested scene in the supplemental material.
}
\label{table:latencybreakdownallprobes}
\vspace{-0.5cm}
\end{table}

\paragraph*{Amortized Rendering}
Modern multiplayer games range in player count from pairs (1:1) to hundreds (MMOs) of players sharing a single world that could be as small as a single room, or as large as a modern city.
%
%
The more players that share a common set of probes, the more a server can amortize the computation affiliated with updating those probes.

Diffuse GI data (in the form of probe \irradiance\ and \visibility) is view independent and is globally rendered by the server for all clients once per update. As a result any cost affiliated with this global update can be effectively amortized across clients.
However, any client-view specific computation cannot be easily amortized. For example, if client view-based culling is used to reduce probe count, the remaining pipeline must be completed on a per-client basis.
Our approach is at its best when many players share visibility of a common volume that, once updated, can be re-used many times across players. As scenes grow larger, alternate techniques such as per-player probe volumes can effectively reduce server workload at the cost of reduced probe data re-use amortization.

%% file: sections/discussion.tex
\section{Discussion}\label{sec:discussion}

\paragraph*{Visibility Compression}
The visibility information in irradiance probes must be compressed losslessly to ensure preventing of light and shadow leaks. Our use of hardware encoding and delta compression is effective but does not take full advantage of spatial entropy analysis as in supercompression methods. A next step for future work is to combine our method with other techniques developed for depth compression (e.g., \cite{lindstrom2014fixed,collet2018zstandard}) to improve compression of distance-based terms for indirect lighting.

When using lossy compression, splitting 16-bit floating-point values across YUV color channels is a poor strategy, as a potential loss of information in bits at arbitrary positions in the values merged from multiple color channels can create severe artifacts in the reconstructed texture data.
An alternative is quantization to a lower range of integer values so that the values can be expressed in a single color channel. Although not without artifacts this approach limits the severity of potentially visible artifacts in rendering.

Certain software implementations of H.264 and HEVC support higher bit depth encoding (12 to 16 bit). However, hardware encoding and decoding of 16-bit values is not yet available on consumer GPUs.

As an alternative, research by Liu \etal introduces a combination of lossless compression of the most significant bits and lossy compression otherwise \cite{liu2015hybrid}. This strategy can improve the compression ratio but increases transcoding complexity for servers and clients.
We leave such investigations for lossy \visibility\ compression to future work.

\paragraph*{Color Compression}
Our work focuses on lossless probe \irradiance\ compression for maximum image quality. As our compression analysis in Section \ref{sec:texturecompression} shows, to some extent probes contain redundancy along the angle (probe texel), position, and time dimensions. Lossy compression could leverage this redundancy more successfully to further reduce the required bandwidth.
Careful analysis of the perceptual impact of compression artifacts in GI data is needed to exploit lossy compression. This is left to future work.

\paragraph*{High Frequency Glossy Surfaces}
Our system handles GI on rough glossy and diffuse surfaces by sending frequency-limited incident lighting data, and does so with correct occlusion. This is an improvement over previous systems limited to only diffuse or direct illumination that did not model dynamic occlusion. 
However, our approach does not simulate high-frequency glossy effects for smooth glossy surfaces and mirror reflections as the angular resolution of the probes is too low. Increasing the probe resolution directly would compromise their efficiency.

\begin{figure}[t]
\centering
\includegraphics[width=\linewidth]{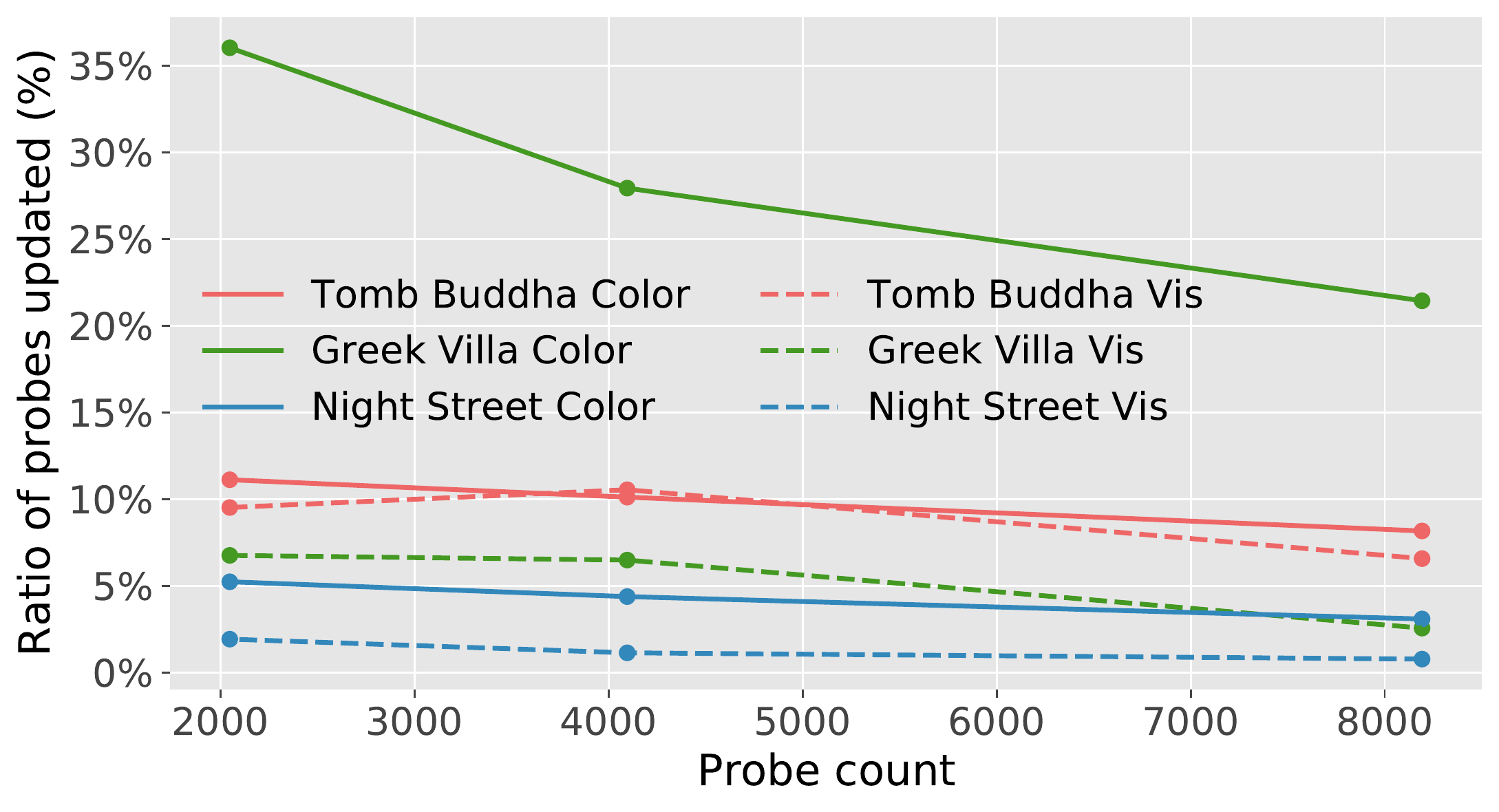}
\vspace{-0.3cm}
\caption{\textbf{Selective Probe Reduction Performance.}~ \normalfont  Average percentage of probes updated (over time) with our potentially visible set method and the unchanged probes removed. Results are shown across various scenes and probe counts. We achieve reduction in transmitted probe count of between 65\% and 99.5\% of active probes. See the supplemental material for further comparison.}
\label{fig:SelectiveProbeUpdatePerformance}
\vspace{-0.5cm}    
\end{figure}

\paragraph*{Variable GI Update Rates}
Depending on the nature of scene dynamics and probe volume organization, it may be possible to update GI for different parts of a scene at rates tuned to their local dynamics.
The simplest version of this approach would involve the use of multiple GI probe volumes, with each volume updated at a rate specific to its coarseness or level of change.
As an example, a small volume, that is camera-locked to the client might be updated at a high rate, while a more coarse volume, capturing global lighting could be updated more slowly.
A more sophisticated approach would entail interleaving probes from each of the volumes described above and combining both such volumes into a single transport based on how much they have changed and how long it has been since their last update for the client.
This approach allows for better utilization of the bandwidth between the client and server to reduce latency for sudden, dramatic scene changes near a viewpoint (such as explosions), while de-prioritizing large-scale global changes (e.g. sun motion) which are much less latency sensitive.

\paragraph*{Latency of Encoding}
Video codec B-frames are relative to both past and future P-frames. They can improve HEVC compression performance by 25\%~\cite{sullivan2012overview}. We prevented the encoder from using B-frames because they inherently add multiple P-frames of delay, in order to look into the future.
However, if the available network throughput is low and transporting a frame with less compression takes longer than multiple smaller frames, then B-frames may be beneficial. 

Furthermore, in scenarios where future game states are predictable, such as architectural walk-throughs or games with slow-moving or deterministic objects, the GI could be made available without delay, or even with look-ahead.

\section{Conclusion}
Distributed graphics is on the rise. The questions of how to distribute the work for interactive systems and where latency is a key consideration are core to its realization. Decoupling clients, pixels, and frames allows better scaling, but introduces system complexity. Early adopters in the distributed interactive graphics market have wisely chosen to use brute force remote rendering and video streaming to minimize complexity for first-generation systems. For the coming generations, more sophisticated approaches with elements such as the ones we have explored may reduce latency and improve end-to-end performance. Shifting the costs of power and computation between client and server is interesting, but reducing the costs on both ends while also improving quality by amortizing work will be the long-term key to creating economically viable and sustainable distributed graphics systems.